# Efficient Control of the Rashba Effective Magnetic Field Using Acceptor-Doped Quantum Wells

Daichi Yamamoto and Takuma Tsuchiya*

*Division of Applied Physics, Faculty of Engineering, Hokkaido University, Sapporo 060-8628, Japan*

To induce a strong Rashba effective magnetic field and enhance its sensitivity to an external electric field, we propose acceptor doping in quantum wells. The acceptors are doped at the center of the well, and donors are doped in the barrier layers to compensate the acceptors and induce conduction electrons. In strongly doped wells, the electric field on these conduction electrons is easily changed by a weak external electric field, by virtue of the strong internal electric field between the acceptors and donors and the resulting high triangle potential barrier induced in the well. As a result, the Rashba effective magnetic field, proportional to the electric field on the electrons, is quite sensitive to the external electric field. Numerical calculations demonstrate that the sensitivity of the Rashba field is larger by two orders of magnitude than that in undoped wells.

1. **Introduction**

The modulation of the Rashba coefficient $\alpha$ [1] using the external electric field $E_{ext}$ is one of the most important key concepts in semiconductor spintronics.[2] In the spin field-effect transistors (spin FETs) proposed by Datta and Das in 1990,[3] for example, the source-drain current is controlled by $E_{ext}$ through the modulation of $\alpha$. Hence, it is desirable that $\alpha$ is sensitive to $E_{ext}$. However, the sensitivity of $\alpha$ to $E_{ext}$ is limited by material parameters. Actually, $\alpha$ is given, for example, by

$$\alpha = \alpha_0 E_{ext},$$
$$\alpha_0 = \frac{\hbar^2}{2m^*} \frac{\Delta}{E_g} \frac{2E_g + \Delta}{(E_g + \Delta)(3E_g + 2\Delta)}, \quad (1)$$

where $\alpha_0$ is the coefficient of the sensitivity given by the material parameters, $m^*$ the effective mass of a conduction electron, $E_g$ the energy gap, and $\Delta$ the spin split-off energy.[4] The effective Zeeman energy due to the Rashba effective magnetic field for up-spin ($\uparrow$) and down-spin ($\downarrow$) electrons is given by

$$E_R^s = \begin{cases} +\alpha k_\parallel & s = \uparrow \\ -\alpha k_\parallel & s = \downarrow \end{cases}, \quad (2)$$

where $k_\parallel = |\vec{k}| = |(k_x, k_y)|$ is the two-dimensional wave number.

Recently, Gvozdić and Ekenberg[5-7] have pointed out that $\alpha$ is enhanced in wider modulation-doped quantum wells, because of the internal electric field induced between conduction electrons in the well and ionized donors in the barrier layers. They have demonstrated by numerical calculations that $\alpha$ is enhanced by one order of magnitude in an 80 nm InGaSb quantum well compared with that expected from $E_{ext}$. On the other hand, Koga and colleagues[8,9] proposed an *n-p-n* triple-barrier structure for a spin-filter device. In this structure, the central barrier and two other outer barriers are *p*- and *n*-type, respectively, and the symmetric internal electric field between the *p*- and *n*-type barriers induces a strong Rashba effective field.

In the present study, we propose to integrate the above two concepts and consider to dope acceptor impurities at the center of the well. Donors are also doped in the barrier layers to compensate the acceptors and induce conduction electrons. Owing to the strong electric field between ionized acceptors and donors, $\alpha$ is expected to be quite sensitive to $E_{\text{ext}}$ even in narrower wells.

This paper is organized as follows. In Sect. 2, we explain the mechanism of the enhancement of $\alpha$ by acceptor doping. In Sect. 3, a method of numerical calculations is shown, and numerical results are presented in Sect. 4. The results are discussed and summarized in Sects. 5 and 6.

## 2. Acceptor Doping in Quantum Wells

Let us consider a quantum well with acceptor impurities doped in it and donors in the barrier layers. We assume here, for simplicity, that the acceptors with a sheet density $N_{\text{A}}$ are $\delta$-doped at $z = 0$, or at the center of the well, and donors with a total sheet density $N_{\text{D}}(> N_{\text{A}})$ are also $\delta$-doped symmetrically in the barrier layers, as shown in Fig. 1(a). This doping profile is similar to that for a spin-filtering device proposed in Refs. 8 and 9.

Since $N_{\text{D}} > N_{\text{A}}$, the acceptors are compensated completely, and conduction electrons with a sheet density $N_{\text{S}} = N_{\text{D}} - N_{\text{A}}$ are induced in the well. Because of this compensation, the acceptors and donors are ionized and a local electric field $\pm E_{\text{DA}}$ is induced between them. As a result, a triangle potential barrier emerges in the well. For a sufficiently large $N_A$, the wave function $\varphi_0(z)$ of the ground subband is strongly modified, and the energy splitting $\Delta E_{sb}^{10}$ between the ground and first excited subbands is small.

The mechanism of the modulation of the Rashba field is essentially the same as those described in Refs. 5-7. We show a schematic illustration of the ground-subband wave function in Fig. 1. For $E_{ext} = 0$ [Fig. 1(a)], the ground-subband wave function is spatially symmetric. Then, the average electric field $E_{av}$ on an electron vanishes, despite the strong local electric field. With increasing $E_{ext}$ [see Fig. 1 (b)], the ground-subband wave function localizes on the lower-energy or the left-hand side in the well, and $E_{av}$ on an electron approaches $E_{DA}$. If $E_{ext}$ necessary for the electron localization is much lower than $E_{DA}$, $\alpha$ is expected to be much higher than that corresponding to $E_{ext}$. In contrast, the first excited state localizes on the right-hand side. Hence, $E_{av}$ on a first excited electrons approaches $-E_{DA}$, and the sign of $\alpha$ becomes opposite to that for the ground-state one.

Note that the present mechanism to induce the Rashba field is different from the conventional one, in which $\alpha$ is induced directly by $E_{ext}$ in accordance with Eq. (1). In the present mechanism, the role of $E_{ext}$ is not to change the electric field on an electron, but to change the location of an electron by modulating the wave function. The Rashba field is induced by the built-in local electric field $\pm E_{DA}$ on localized electrons.[5,6,10,11]

Using a two-level tight binding model,[12] we can discuss the $E_{ext}$ dependence of $\alpha(\propto E_{av})$ semiquantitatively under the condition of a high $N_A$ and a low $E_{ext}$. Consider two electronic states localized on the left and right sides of the well. Their normalized wave functions are $\phi_L(z)$ and $\phi_R(z)$, and their energies are $\varepsilon_L$ and $\varepsilon_R$, respectively.

The wave functions of the ground and first excited subbands are approximated by $\varphi_{n(=0,1)}(z) = a_{nL}\phi_L(z) + a_{nR}\phi_R(z)$, where $n=0$ and 1 denote the ground and first excited subbands, respectively, and the coefficients $a_{nL}$ and $a_{nR}$ are normalized as $a_{nR}^2 + a_{nL}^2 = 1$. The matrix equation for this system is

$$\begin{pmatrix} (\varepsilon_L - \varepsilon_n) & -V \\ -V & (\varepsilon_R - \varepsilon_n) \end{pmatrix} \begin{pmatrix} a_{nL} \\ a_{nR} \end{pmatrix} = 0. \tag{3}$$

Here, $\varepsilon_n$ denotes the eigenenergy, and the coupling $V$ is given by $V = \Delta E_{sb}^{10}/2$, where $\Delta E_{sb}^{10} = \varepsilon_1 - \varepsilon_0$ is the subband splitting for $\varepsilon_R = \varepsilon_L$. Solving Eq. (3), we obtain

$$\begin{aligned} \varepsilon_0 &= \frac{\varepsilon_R + \varepsilon_L}{2} - \sqrt{V^2 + \Delta^2}, \\ \varepsilon_1 &= \frac{\varepsilon_R + \varepsilon_L}{2} + \sqrt{V^2 + \Delta^2}, \end{aligned} \tag{4}$$

where $\Delta = \Delta\varepsilon_{RL}/2 = (\varepsilon_R - \varepsilon_L)/2$, and the coefficients $a$'s are

$$\begin{aligned} a_{nL} &= \frac{V}{\sqrt{2\left(V^2 + \Delta^2 - (-1)^n \Delta\sqrt{V^2+\Delta^2}\right)}}, \\ a_{nR} &= \frac{-\Delta + (-1)^n \sqrt{V^2 + \Delta^2}}{\sqrt{2\left(V^2 + \Delta^2 - (-1)^n \Delta\sqrt{V^2+\Delta^2}\right)}}, \end{aligned} \tag{5}$$

Then, $E_{av}$ is given by

$$E_{av,n} = a_{nL}^2 E_L + a_{nR}^2 E_R, \tag{6}$$

where $E_L = E_{DA} + E_{ext}$ and $E_R = -E_{DA} + E_{ext}$ are the local electric field in the left- and right-hand sides of the well, and $\alpha_n = \alpha_0 E_{av,n}$. To estimate $\varepsilon_L$ and $\varepsilon_R$, we need to take into account the energy shift due to $E_{ext}$ and the local Rashba field $\pm \alpha_0 E_{L(R)} k_\parallel$ as

$$\begin{cases} \varepsilon_L = -eE_{ext}\Delta z_{RL}/2 - \alpha_0 E_L k_\| \\ \varepsilon_R = eE_{ext}\Delta z_{RL}/2 + \alpha_0 E_R k_\| \end{cases} \quad (7)$$

for up-spin states, and

$$\begin{cases} \varepsilon_L = -eE_{ext}\Delta z_{RL}/2 + \alpha_0 E_L k_\| \\ \varepsilon_R = eE_{ext}\Delta z_{RL}/2 - \alpha_0 E_R k_\| \end{cases} \quad (8)$$

for down-spin states, where $\Delta z_{RL}$ is the distance between the average positions of $\phi_L(z)$ and $\phi_R(z)$. We obtain the energies of up-spin states, $E_{0\uparrow}(k_\|)$ and $E_{1\uparrow}(k_\|)$, by solving Eq. (3) with Eq. (7), and those of down-spin states, $E_{0\downarrow}(k_\|)$ and $E_{1\downarrow}(k_\|)$, by solving Eq. (3) with Eq. (8). The Rashba spin splitting is given by

$$\Delta E_R^n(k_\|) = E_{n\uparrow}(k_\|) - E_{n\downarrow}(k_\|). \quad (9)$$

Let us consider the behavior of $\Delta E_R^n(k_\|)$ through this tight-binding model. For $k_\| = 0$ and $E_{ext} = 0$, the system is spatially symmetric and $\varepsilon_R = \varepsilon_L$ for both spin states. Because $\Delta = 0$ and $E_L = -E_R = E_{DA}$, we obtain $a_R^2 = a_L^2 = 1/2$ and $\alpha = E_{av} = 0$ for up-spin and down-spin states of the ground and first excited subbands, and $\Delta E_R^n = 0$. For $k_\| \neq 0$, the Rashba terms $\pm \alpha E_{L(R)} k_\|$ in Eqs. (7) and (8) cause the difference between $a_L^2$ and $a_R^2$, which means the localizations of the up-spin and down-spin states on opposite sides. [This behavior is qualitatively the same as that shown in Fig. 4(b1).] Thus, each state has a Rashba energy. However, the spin splitting does not occur, because the energy shifts are equal for both spin states. For spin splitting, the breaking of the spatial symmetry caused by $E_{ext}$ is necessary. When $\Delta \gg V$, $a_{0L} \simeq 1$ and $a_{0R} \simeq 0$ for the ground subband, and $a_{1L} \simeq 0$ and $a_{1R} \simeq 1$ for the first excited one. Thus, $E_{av,0} \simeq E_L = E_{DA} + E_{ext}$ and $E_{\text{av},1} \simeq E_{\text{R}} = -E_{\text{DA}} + E_{\text{ext}}$ for the ground and first excited subbands, respectively. Since $V$ can be quite small for a high $N_{\text{A}}$, the above condition is easily satisfied even for a small $E_{\text{ext}}$. Then, $E_{\text{av}}$ is much larger than $E_{\text{ext}}$ for $|E_{\text{DA}}| \gg |E_{\text{ext}}|$, and the Rashba field is expected to be quite sensitive to $E_{\text{ext}}$.

## 3. Numerical Method

To demonstrate the validity of the present mechanism, we perform numerical calculations. Since the electric field due to conduction electrons affects the electronic states to some extent, we need to solve the Schrödinger and Poisson equations self-consistently to obtain reliable results. The self-consistent potential is important particularly for small-$N_{\text{A}}$ samples. For large-$N_{\text{A}}$ samples, the tight-binding model gives good results, as will be shown in Sect. 5.

We employ the effective mass approximation for electronic states for simplicity. This simple approach neglects some effects, such as the intersubband-induced spin-orbit interaction recently predicted in Refs. 10 and 11. However, the present approach is sufficient for the purpose of this study.

Considering the translational symmetry along the $xy$-direction, we can separate the Schrödinger equation into the $z$- and $xy$-parts, and the $z$-part is given by

$$-\frac{\hbar^2}{2m^*}\frac{d^2\varphi_{n\uparrow(\downarrow)}(z,k_\parallel)}{dz^2} + \left[V_{\text{b}}(z) + eE_{\text{ext}}z + V_{\text{e}}(z) \pm E_{\text{R}}^{n\uparrow(\downarrow)}(z,k_\parallel)\right]\varphi_{n\uparrow(\downarrow)}(z,k_\parallel) \quad (10)$$
$$= E_{n\uparrow(\downarrow)}(k_\parallel)\varphi_{n\uparrow(\downarrow)}(z,k_\parallel),$$

where $\varphi_{n\uparrow(\downarrow)}$ and $E_{n\uparrow(\downarrow)}$ are respectively the normalized wave function and eigenenergy along the $z$-direction for a up-spin ($\uparrow$) or down-spin ($\downarrow$) conduction electron in the $n$th subband, $V_b(z) = \Delta V_b[\Theta(z-d/2)+\Theta(-z+d/2)]$ the barrier potential, with $\Theta$ the Heaviside step function and $d$ the well thickness, $V_e(z)$ the electronic potential energy comes from conduction electrons, donors, and acceptors, and $E_R^{n\uparrow(\downarrow)}(z,k_\parallel)$ the local effective Zeeman energy comes from the Rashba effective magnetic field. Because of the rotational symmetry of the Rashba field, $\varphi_{n\uparrow(\downarrow)}$ and $E_{n\uparrow(\downarrow)}$ are independent of the direction of $\vec{k}_\parallel$. When we consider the $xy$-direction, the wave function and eigenenergy are given respectively by

$$\Phi_{n\uparrow(\downarrow)}(\mathbf{r},k_\parallel) = \varphi_{n\uparrow(\downarrow)}(z,k_\parallel)\frac{\exp[i(k_x x + k_y y)]}{L}, \tag{11}$$

and

$$\varepsilon_{n\uparrow(\downarrow)}(k_\parallel) = \frac{\hbar^2 k_\parallel^2}{2m^*} + E_{n\uparrow(\downarrow)}(k_\parallel), \tag{12}$$

where $L$ is the sample dimension along the $x$- and $y$-direction. The effective Zeeman energy $E_R^{n\uparrow(\downarrow)}(z,k_\parallel)$ is given, according to Eq. (2), by

$$E_R^{n\uparrow(\downarrow)}(z,k_\parallel) = \pm\alpha_0 E_e(z)k_\parallel,$$
$$E_e(z) = \frac{1}{e}\frac{d}{dz}V_e(z) + E_{ext}, \tag{13}$$

where the "$+$" sign is for a up-spin state and the "$-$" is for a down-spin state. Note that Eq. (13) underestimates the Rashba field, because the effects of the heterointerface are not taken into account.[7,13] The electronic potential $V_e(z)$ is obtained by solving the Poisson equation

$$\frac{d^2V_e(z)}{dz^2} = -\frac{dE_e(z)}{dz} = -\frac{e}{\varepsilon}\left[-n_e(z) - n_A(z) + n_D(z)\right],$$

$$n_e(z) = \sum_n \sum_{\uparrow,\downarrow} \int_0^{k_F^{n\uparrow(\downarrow)}} \frac{1}{(2\pi)^2} \left|\varphi_{n\uparrow(\downarrow)}(z, k_\parallel)\right|^2 2\pi k_\parallel dk_\parallel, \quad (14)$$

$$n_A(z) = N_A \delta(z),$$

$$n_D(z) = \frac{N_D}{2}\left\{\delta\left[z - \left(\frac{d}{2} + \delta_S\right)\right] + \delta\left[z + \left(\frac{d}{2} + \delta_S\right)\right]\right\},$$

where $\varepsilon$ is the dielectric constant, $k_F^{n\uparrow(\downarrow)}$ the Fermi energy for up-spin and down-spin electrons in the $n$th subband, $n_e$, $n_A$, and $n_D$ the densities of electrons, acceptors, and donors, respectively, and $\delta_S$ the thickness of the spacer layer.

In actual numerical calculations, we expand all $z$-dependent quantities, such as $\varphi_{n\uparrow(\downarrow)}$, $E_R^{n\uparrow(\downarrow)}$, $V_e$, $V_b$, and $E_e(z)$, into a Fourier series. In iteration, we use the so-called attenuated mixing to avoid instability.[14] The new input potential $V_e$ for Eq. (10) is a mixture of 30% of the new output $V_e$ of Eq. (14) and 70% of the old input $V_e$ of Eq. (10). After 35 iterations, the change in eigenenergies in an iteration step is below $10^{-5}$ meV.

## 4. Numerical Results

In Fig. 2, we show the dispersion relation $\varepsilon_{n\uparrow(\downarrow)}(k_\parallel)$ of the ground ($n=0$) and first excited ($n=1$) subbands in a 17 nm $In_{0.53}Ga_{0.47}As/Al_{0.52}Ga_{0.48}As$ quantum well with $N_A = 12\times10^{12}$ cm$^{-2}$, $N_S = 1\times10^{12}$ cm$^{-2}$, $N_D = 13\times10^{12}$ cm$^{-2}$, and $\delta_S = 1$ nm under $E_{ext} = 10^6$ V/m. Note that such high densities of $\delta$-doped acceptors and donors, $N_A$ and $N_D$, respectively, are experimentally possible in III-V semiconductors.[15] The material parameters used in the present calculations are summarized in Table I. For simplicity, we ignore the difference in $\varepsilon$ and $m^*$ between the well and barrier layers. These dispersion relations show clear Rashba spin splitting $\Delta E_R^n(k_\|)$. For example, $\Delta E_R^0(k_{F0}) = 7.0$ meV at $k_{F0} \simeq 0.194$ nm$^{-1}$ or the Fermi wavelength for the ground subband. However, for $N_A = 0$, or for a quantum well without acceptors, the dispersion relations for the up-spin and down-spin states are almost degenerate and $\Delta E_R$ is only 0.1 meV. Thus, the Rashba field is enhanced by almost two orders of magnitude by acceptor doping. For the first excited subband, $\Delta E_R^1(k_{F0}) = -6.8$ meV $\simeq -\Delta E_R^0(k_{F0})$.

In Fig. 3(a), $\Delta E_R^n(k_\|)$ at $k_\| = 0.2$ nm$^{-1}$ is shown as a function of $E_{ext}$ for some values of $N_A$. It is clear that the Rashba coefficient $\alpha (= \Delta E_R/2k_\|)$ depends on $N_A$ and is quite sensitive to $E_{ext}$ for a larger $N_A$. The signs of $\Delta E_R^n$ are different between the ground and first excited subbands, except for $N_A = 0$. For $N_A = 12 \times 10^{12}$ cm$^{-2}$, for example, $\Delta E_R^n$ increases with $E_{ext}$ quite rapidly for $E_{ext} < 0.7 \times 10^6$ V/m and saturates for a larger $E_{ext}$. The factor of the enhancement $\Delta E_R^n(E_{ext}, N_A)/\Delta E_R^0(E_{ext}, N_A = 0)$ is shown in Fig. 3(b). This factor increases with $N_A$ and reaches 100 for $N_A = 12 \times 10^{12}$ cm$^{-2}$ for $E_{ext} < 0.5 \times 10^6$ V/m.

To confirm the mechanism of the modulation of $\alpha$ by the wave function modulation, we show in Fig. 4 the potential profile and wave functions of the ground and first excited subbands for $N_A = 12 \times 10^{12}$ cm$^{-2}$. Figure 4(a1) shows the results for $k_\| = 0$ and $E_{ext} = 0$. The wave functions of the up-spin and down-spin states are the same, because there is no Rashba field for $k_\| = 0$. The energy splitting between the first excited and ground subbands is $\Delta E_{\text{sb}}^{10} = \varepsilon_{1\uparrow(\downarrow)}(0) - \varepsilon_{0\uparrow(\downarrow)}(0) = 1.83$ meV. The wave function of the ground subband is spatially symmetric and that of the first excited one is antisymmetric. Then, the probability amplitudes are symmetric, and the average electric field on an electron and the Rashba coefficient vanish for both states. In Fig. 4(a2), we show the results for $E_{\text{ext}} = 10^6$ V/m. The wave functions of the ground and first excited subbands are localized on the left- and right-hand sides almost completely. This is because the energy difference comes from $E_{\text{ext}}$ between the left and right interfaces, $\Delta \varepsilon_{\text{RL}} \simeq 14$ meV, exceeds $\Delta E_{\text{sb}}^{10} = 1.83$ meV. Because of this localization, the average electric fields on electrons in the ground and first excited subbands are about $E_{\text{DA}}$ and $-E_{\text{DA}}$, respectively. These electric fields are much stronger than $E_{\text{ext}}$ for a larger $N_{\text{A}}$, and the resulting Rashba field is expected for a finite $k_{\parallel}$ to be much stronger than that expected from $E_{\text{ext}}$.

Wave functions are modified for $k_{\parallel} \neq 0$ by the local Rashba field $\pm \alpha_0 E_{\text{e}}(z) k_{\parallel}$ even when $E_{\text{ext}} = 0$.[5,6,8)] Numerical results are shown in Fig. 4(b1) for $E_{\text{ext}} = 0$ and $k_{\parallel} = 0.2$ nm$^{-1}$. Because of the local Rashba field induced by the internal electric field $\pm E_{\text{DA}}$, the energies are lower on the left and right sides of the well for the down-spin and up-spin states, respectively. As a result, the up-spin and down-spin electrons in the ground subband localized on the left- and right-hand sides, and those in the first excited subband on the opposite side. Despite this spatial separation between the up-spin and down-spin states, there is no spin splitting in each subband. This is because the energy shifts due to the local Rashba field are equal for both spin states. This spatial spin separation vanishes under sufficient $E_{\text{ext}}$, as shown in Fig. 4(b2).

In Fig. 5, we show the wave functions for $N_{\text{A}} = 0$ for comparison. Even under $E_{ext} = 10^6$ V/m, the wave functions are modified only slightly and the average electric field for both subbands is about $E_{ext}$.

## 5. Discussion

The sensitivity of $\alpha$ to $E_{ext}$ of the acceptor-doped wells, shown in our numerical calculations, is quite desirable for device applications. For spin FETs,[3] for example, the transconductance $g_m = dI_{DS}/dV_{GS}$, one of the most important device parameters for FETs, is expected to improve by two orders of magnitude, where $I_{DS}$ is the electric current between the drain and source electrodes and $V_{GS} (\propto E_{ext})$ is the voltage between the gate and the source. Then, the maximum operation frequency, proportional to $g_m$,[16] is also expected to improve. Furthermore, $\Delta V_G$, or the difference in $V_{GS}$ necessary for the switching of $I_{DS}$, is also expected to decrease by two orders of magnitude. Then, the power consumption, which is necessary for the charge and discharge of the gate capacitance and the stray capacitance of electric circuits and is proportional to $\Delta V_G^2$, is expected to improve by four orders of magnitude at most.

The localization of wave functions shown in Fig. 4(b1) seems to indicate a possibility of the spatial separation of the up-spin and down-spin electrons in the ground subband. However, it is actually not the case, because the directions of the up and down spins depend on the direction of $\vec{k}_\parallel$. Then, the focusing of the direction of $\vec{k}_\parallel$ is necessary for the spatial spin separation, and it may be realized, for example, by current flow in a quasi-one-dimensional channel. The extraction of spin-polarized electrons on one side of double-well structures is technically possible.[17-20]

The present structure has some possible deficiencies. One of them is the electron population in the first excited subband shown in Fig. 2. This is caused mainly by the lowering of the subband splitting due to the triangular potential in the well, and causes the mixture of electrons with the opposite Rashba field. We can avoid this problem by decreasing $N_A$ and $N_S$, although the sensitivity of $\alpha$ to $E_{ext}$ is also decreases to some extent. However, the electron population in excited subbands is not always an obstacle. Consider the transport of spin-polarized electrons injected from a source electrode. The directions of the precession, caused by the Rashba field, are opposite between electrons in the ground and first excited subbands with the same $\vec{k}_\parallel$. Thus, the spin polarization decreases with spin rotation. However, the frequencies of the spin precession for the electrons in both subbands are almost the same, because $\Delta E^0_{Rashba}(k_\parallel) \simeq -\Delta E^1_{Rashba}(k_\parallel)$. Hence, the spin polarization is expected to be restored at precession angles $|\theta_{pr}| = n\pi$ $(n = 1, 2, 3, \cdots)$. Therefore, the present problem is avoidable by appropriate device operation.

Another possible defect is low electron mobility due to impurity scattering by the high-density acceptors in the well. However, the probability amplitude of electrons is relatively low in the acceptor-doped region, because of the high triangle potential, and the scattering rate is expected to be sufficiently low. In any case, the scattering rate should be estimated theoretically in future studies.

The two-level tight-binding model reproduces the results of the self-consistent calculations quite well for a higher $N_A$. The Rashba spin splitting $\Delta E^n_R$, given by the tight-binding calculation, is shown in Fig. 6 with the self-consistent results shown in Fig. 3(a). The parameters used for this tight-binding estimation, namely, the subband splitting $\Delta E_{sb}^{10}$ and the distance between the left and right peaks of the wave function, $\Delta z_{RL}$, are summarized in Table II. The parameters for $N_A = 12 \times 10^{12}$ (cm$^{-2}$), for example, are obtained from the results shown in Fig. 4(a1). The results of the tight-binding estimation are in accordance with the self-consistent results, and it is clear that the tight-binding model is sufficient for a semiquantitative estimation of $\alpha$ at least for $N_A \geq 4 \times 10^{12}$ cm$^{-2}$ and $E_{ext} \leq 2 \times 10^{12}$ V/m.

Finally, note that it is also possible to control the Rashba splitting in similar symmetric potential structures by applying an external *magnetic* field.[21] Recently, a spin blocker device utilizing the magnetic control has been proposed.[22]

## 6. Summary

To enhance the sensitivity of the Rashba effective magnetic field to an external electric field, we proposed the use of a quantum well structure in which acceptors and donors are doped at the center of the well and in the barriers, respectively. The strong internal electric field between ionized acceptors and donors enhances the sensitivity. Numerical calculations have shown that the Rashba field can be two orders of magnitude more sensitive to the external field than the usual quantum wells without acceptors. We hope that this structure will contribute to the progress of the field of spintronics.

*E-mail: t.t@physics.org

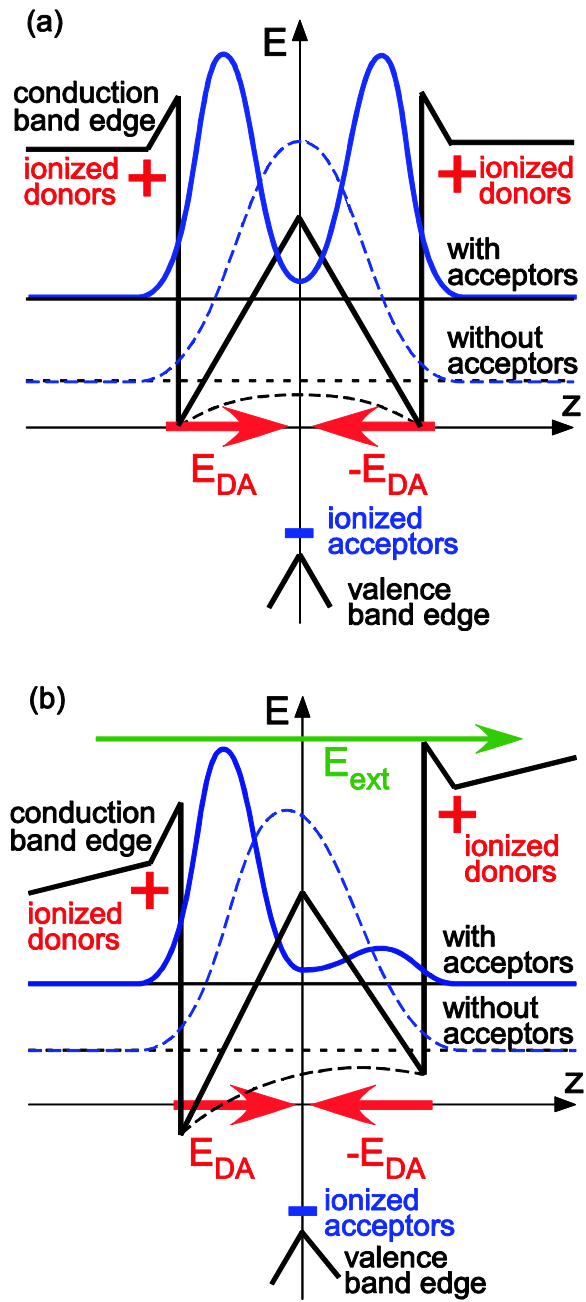

Fig. 1. (Color online) Schematic illustrations of the ground-state wave function and potential profile in an acceptor-doped quantum well (a) without and (b) with external electric field.

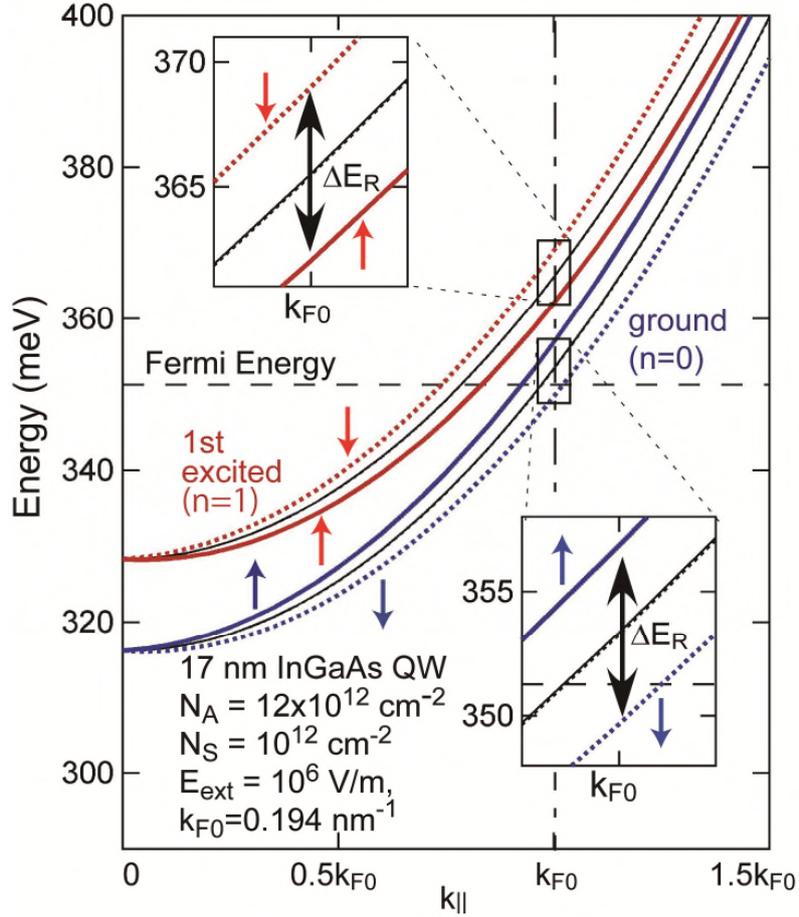

Fig. 2. (Color online) Dispersion relation of the up-spin (solid lines) and down-spin (dotted lines) states of the ground ($n=0$) and first excited ($n=1$) subbands for a 17 nm $In_{0.53}Ga_{0.47}As/Al_{0.52}Ga_{0.48}As$ quantum well with $N_A = 12 \times 10^{12}$ cm$^{-2}$, $N_S = 1 \times 10^{12}$ cm$^{-2}$, $N_D = 13 \times 10^{12}$ cm$^{-2}$, and $\delta_S = 1$ nm under $E_{ext} = 10^6$ V/m. Results for $N_A = 0$, shown by thin black lines, are shifted so that the energies match those for $N_A = 12 \times 10^{12}$ cm$^{-2}$ at $k_\parallel = 0$.

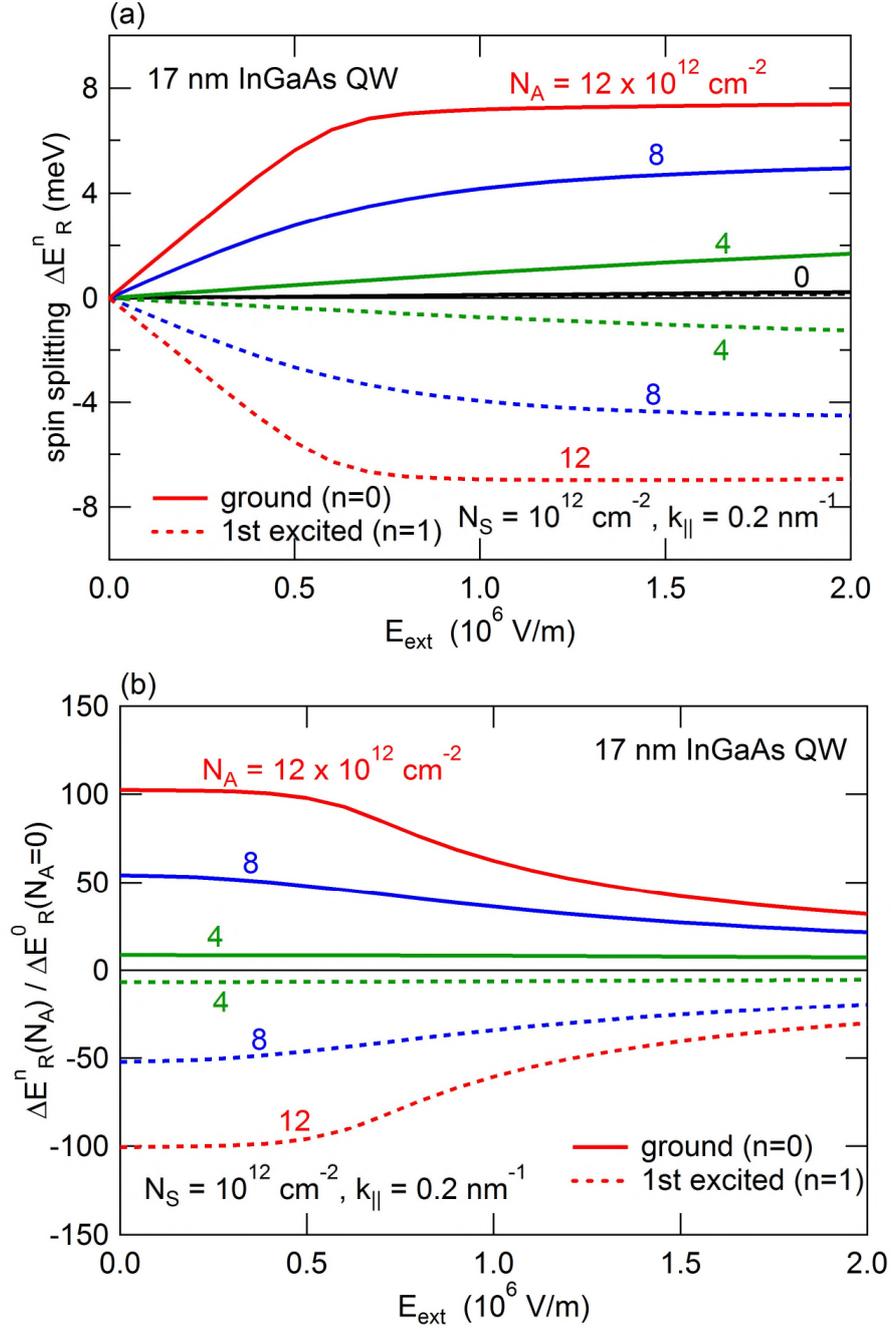

Fig. 3. (Color online) External field dependences of (a) the spin splitting and (b) the enhancement ratio for the ground and first excited subbands in a 17 nm $In_{0.53}Ga_{0.47}As/Al_{0.52}Ga_{0.48}As$ quantum well. The solid and dashed lines show results for the ground ($n=0$) and first excited ($n=1$) subbands, respectively.

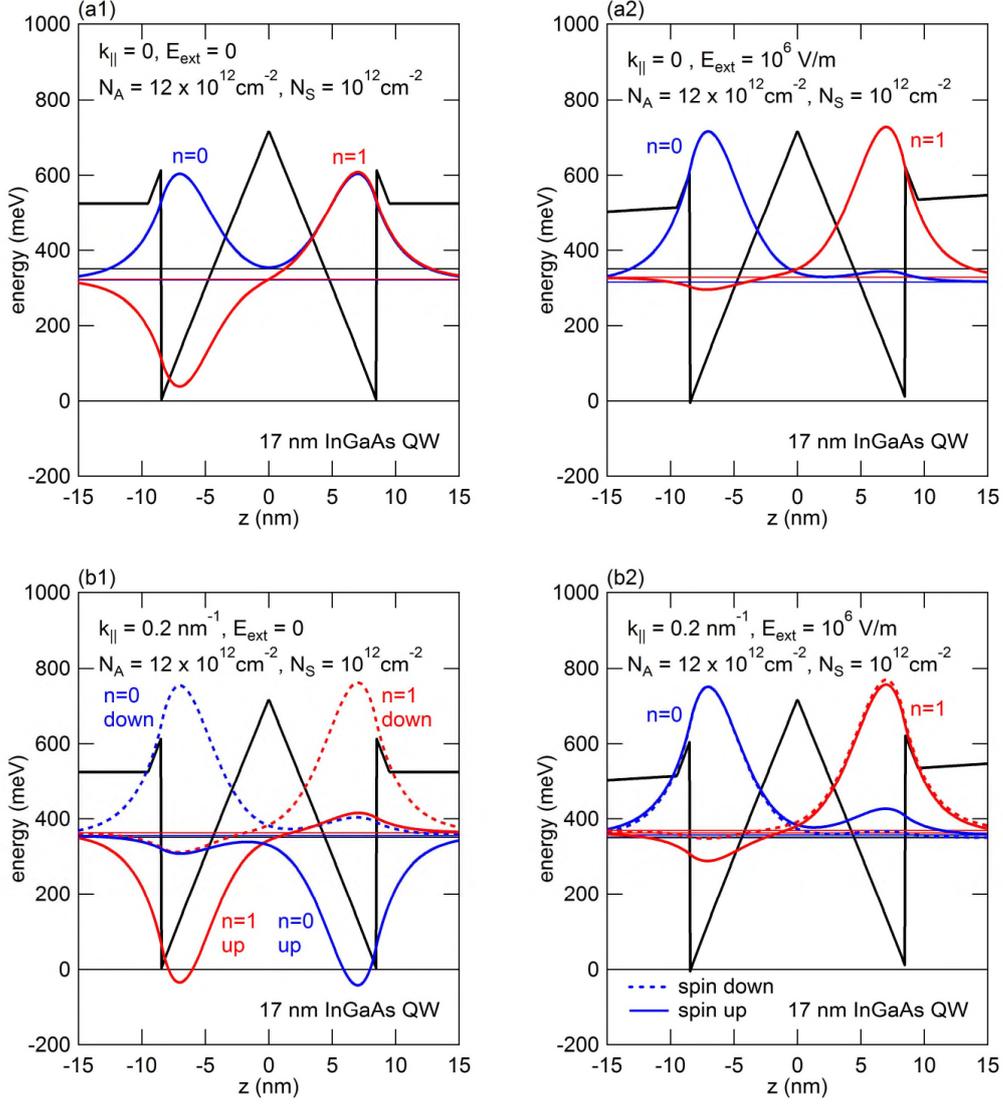

Fig. 4. (Color online) Wave functions of the ground and first excited subbands in a 17 nm In$_{0.53}$Ga$_{0.47}$As/Al$_{0.52}$Ga$_{0.48}$As quantum well with $N_A = 12\times10^{12}$ cm$^{-2}$, $N_S = 1\times10^{12}$ cm$^{-2}$, $N_D = 13\times10^{12}$ cm$^{-2}$, and $\delta_S = 1$ nm.

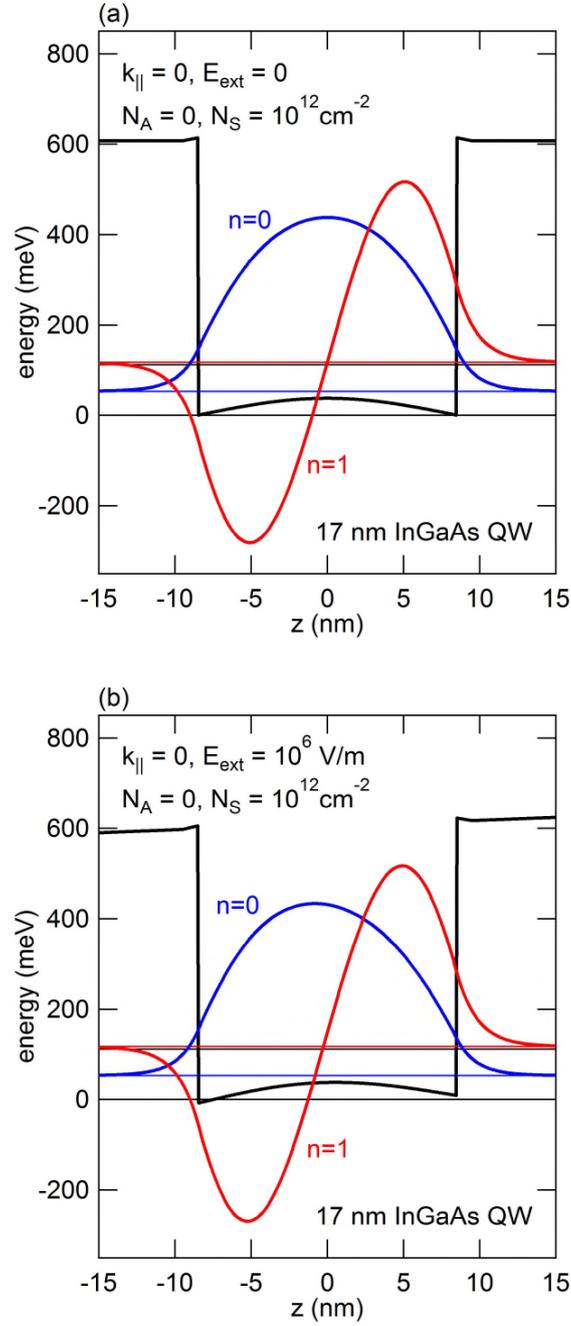

Fig. 5. (Color online) Wave function of the ground and first excited subbands in a 17 nm In$_{0.53}$Ga$_{0.47}$As/Al$_{0.52}$Ga$_{0.48}$As with $N_A = 0$, $N_S = 1\times10^{12}$ cm$^{-2}$, $N_D = 1\times10^{12}$ cm$^{-2}$, and $\delta_S = 1$ nm.

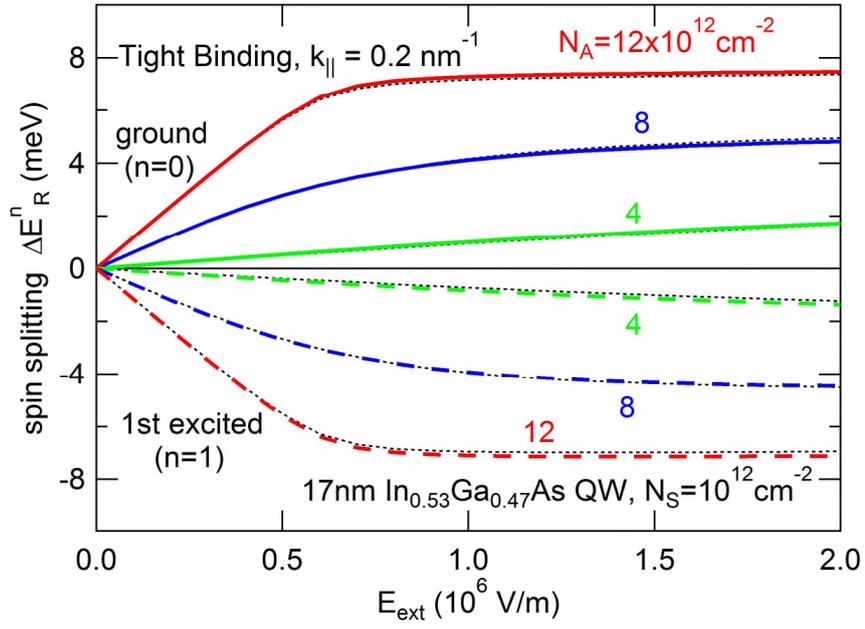

Fig. 6. (Color online) Tight-binding results of external field dependence of the spin splitting in a 17 nm $In_{0.53}Ga_{0.47}As/Al_{0.52}Ga_{0.48}As$ quantum well. The self-consistent results shown in Fig. 3(a) are also shown by thin dotted lines.

Table I. Parameters used in the numerical calculations.[9]

| Material | In$_{0.53}$Ga$_{0.47}$As/Al$_{0.52}$Ga$_{0.48}$As |
|---|---|
| Effective mass $m^*$ | 0.041 $m_0$ |
| Energy gap $E_g$ | 0.783 eV |
| Spin-orbit splitting $\Delta$ | 0.328 eV |
| Dielectric constant $\varepsilon$ | 13.1 |
| Band discontinuity $\Delta V_b$ | 0.614 eV |

$m_0$ : electron rest mass

Table II. Parameters used in the tight-binding calculations.

| $N_A$ ($10^{12}$ cm$^{-2}$) | $\Delta E_{sb}^{10}$ (meV) | $\Delta z_{RL}$ (nm) |
|---|---|---|
| 4 | 24.06 | 9.86 |
| 8 | 6.87 | 10.75 |
| 12 | 1.83 | 12.13 |